# Periodical component found in the space maser signal and its interpretation

Sergey Siparov

State University of civil aviation, St-Petersburg, Russia

sergey@siparov.ru

Vladimir Samodurov

RAS Radio Astronomy Observatory, Pushchino, Russia

sam@prao.ru

The periodical component predicted by the theory of optic-metrical parametric resonance was recently found in the signal of Sep A radio source. If such observations performed independently give similar results, they would provide another evidence of the gravitational waves (GW) existence and the perspectives of the GW astronomy could be discussed.

## 1. Introduction

In this paper we report of the periodic component found recently in the signal of the space radio source Sep A. The background for the search was the prediction of the optic-metrical parametric resonance (OMPR) theory which suggests that if certain conditions are fulfilled, the action of the periodic gravitational waves (GW) source (like a compact binary or a pulsar) on a monochromatic electromagnetic wave (EMW) of a space maser would cause the special type of a parametric resonance, and a specific component would appear in the observed maser signal. This effect has a resonance character, therefore, it is of the zeroth order in the dimensionless amplitude of the GW. This makes it principally different from all the 18 other known effects [1] that could be used for the detection of the GW – all of them are of the first order in the extremely small amplitude of the GW.

## 2. Space maser signal

The monochromatic cosmic radio radiation corresponding to certain transitions in the H$_2$O, OH and other molecules is produced by the space masers [2 – 5] – the clouds with the diameters of $10^0 \div 10^2$ a.u.. Such masers radiation spectra occupy the bands not more than $10^2$ km/s wide and contain the series of narrow details corresponding to the spots with dimensions of about 0.1a.u.. The parts of the maser cloud can have various densities and can move relative to one another. The origin of the pumping may vary for various masers and the spectra and the intensities of their details may also vary with time. In [6 - 8] it was discovered that the space maser radiation can have ultra-rapid fluctuations (tens of minutes) whose origin could be due both to internal processes inside the maser clouds and to some external reasons. Ultra-rapid variability of the radiation can be explained by the competition between the close spots for the radiation pumping; by the changes inside the spots; by the fast superpositions of the spots over the line of sight; by the curl phenomena inside the maser etc. But it seems hard to suggest an internal process leading to the appearance of a periodical component in the intensity of the maser radiation. In some cases it could be explained by an external reason. As it is shown in [11 - 12], at certain conditions the radiation of a space maser can obtain the specific component caused by the action of the periodic GW emitted by a compact binary system or by a pulsar. The theory of such effect - optic-metrical parametric resonance - was developed in [10 -12]. Since the effect is of the zeroth order, the heart of the GW detection problem shifts from the field of the instrumental design and signal/noise ratio to the field of the search for the suitable astrophysical system and its observation with the help of the existing instruments.

The mentioned above specific component presents the periodic change of the intensity of the maser signal with a doubled frequency of the GW proportional to

$$\text{Im}(\rho_{21}) \sim \frac{\alpha_1}{D}\cos 2Dt + O(\varepsilon) \quad (1)$$

Here $\alpha_1 = \mu E/\hbar$ is the so called Rabi parameter (Rabi frequency) proportional to the intensity of the EMW ($\mu$ is the induced momentum, $E$ is the electric stress of the EMW), $D$ is the frequency of the GW, $\varepsilon = \gamma/\alpha_1$ is a small parameter characterizing the strength of the maser field, $\gamma$ -is the decay rate (for the transition in the space H$_2$O-maser $\gamma \sim 10^{-14}$ s$^{-1}$). As it can be seen from eq.(1), the signal in question does not depend on the GW amplitude, and this is characteristic for the resonance effects, its own amplitude depends on the ratio between the Rabi frequency and the GW frequency which is of the order of *O(1)*. Some details of the OMPR theory are given in the Appendix.

## 3. Observations

### *3.1 Maser signals and their processing*

The registration of the space maser signal is performed in the following way. The radio telescope is tuned at a given frequency and its antenna follows the chosen maser for several hours. During a certain exposure time (from 3 to 8 minutes for different sessions and sources) the signal including all the adjacent frequencies is stored, then the spectrum is constructed – the dependence of the flow intensity (Jy ∗ km/s) at a given frequency on the value of the frequency (actually, on the detuning of the given frequency from the main one). Then the exposure regime is switched on again and the procedure repeats $N$ times until the observation conditions permit. As a result, a set of spectra corresponding to the signals received in the subsequent (approximately equal) intervals of time is obtained (Fig.1a-1c).

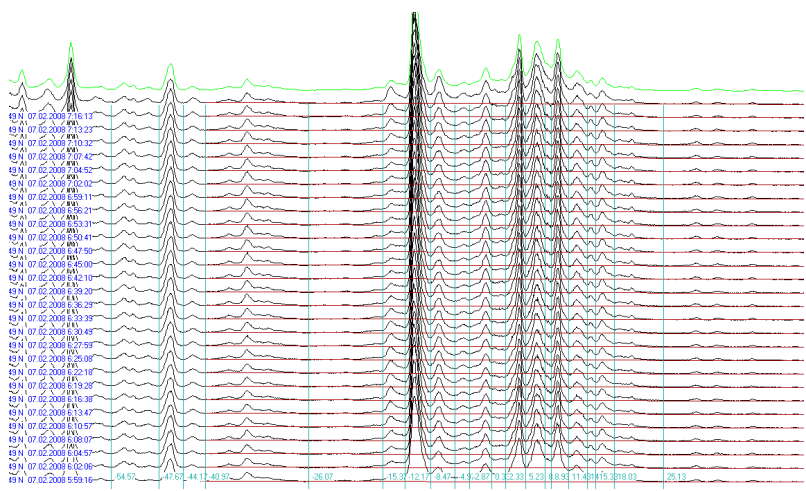

Fig. 1a: W49N spectra, RT-22, Pushchino RAO, 07.02.2008

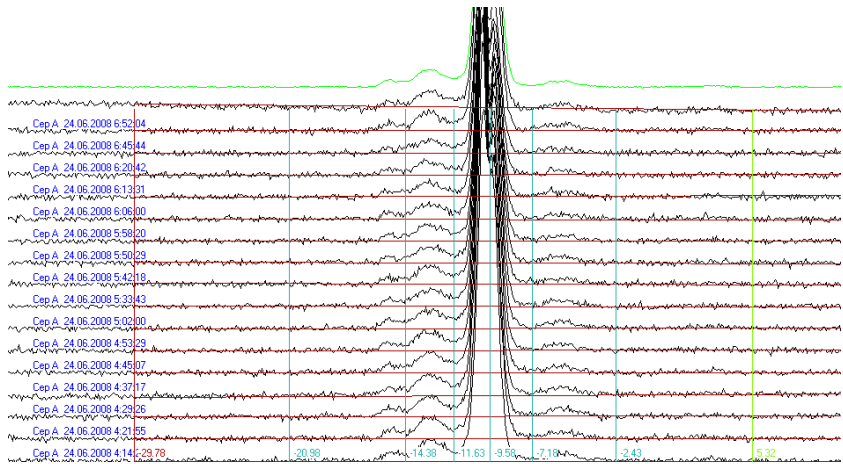

Fig. 1b: W33B spectra, RT-22, Pushchino RAO, 27.06.2008

The processing starts with obtaining the average value of the flow, $I_0$, for every spectrum. Then the interval of frequencies (velocities) is divided into parts containing different details of the spectra (vertical lines on Fig.1), and the time dependent averages, $I_i$, for every part are calculated.

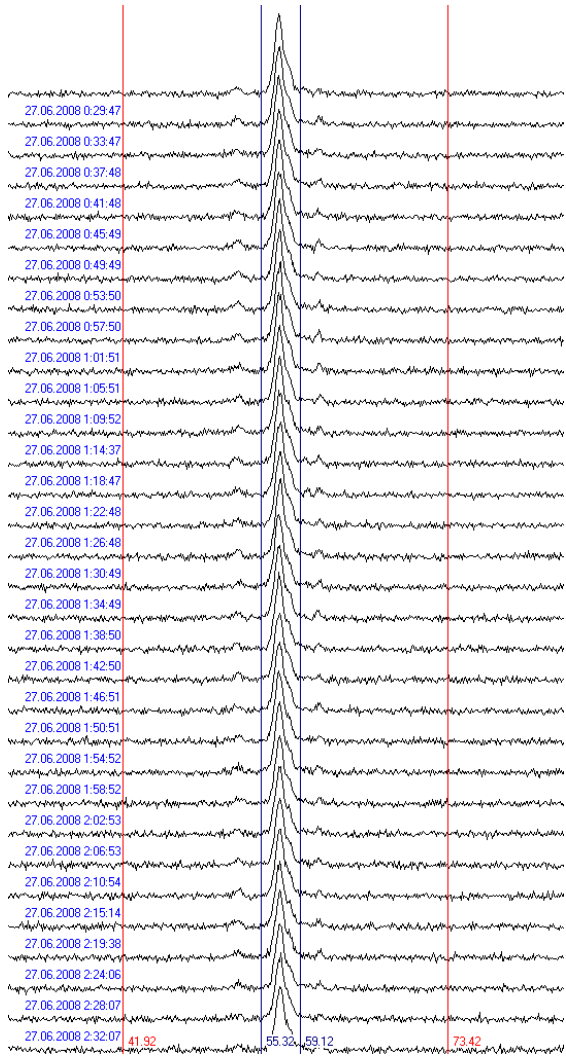

Fig. 1c: Cep A spectra, RT-22, Pushchino RAO, 24.06.2008

Then the differences, $I_{id} = I_i - I_0$, are found, and the software provides the dependences $I_0(n), I_i(n), I_{id}(n)$ for $n = 1, 2, ...N$. The calculation of the differential flows $I_{id}(n)$ makes it possible to exclude the effect of the variability of the source as a whole (and also the effects of the variability of the instrument properties, atmosphere, interstellar medium) on the behavior of the details' variations. If the equidistant time points are needed, the spline approximation can be used.

Then the Fourier transformations for the time dependencies of the flow as a whole and of every chosen detail are performed. If the time dependence of a signal is complicated (stochastic processes), the Fourier transform contains a set of frequencies characterized by the peaks of various heights. In view of our problem, the situation is of interest if the

Fourier transform presents a solitary peak. The presence of such a plot even in one observation session would mean the existence of the periodical (regular) change of the flow for a given spectrum detail at least for the time of observation (several hours). In this case one could define the values of the corresponding frequency, $v_j$, and period $T_j$. Then the corresponding reading of the radio telescope signal gives the intensity $I_{id}^{(j)}$ and the detuning and the frequency of this detail.

Thus, the following values could be obtained as a result of the data processing:

- $v_j$ - the frequency of the periodical flow change of the spectrum detail ;
- $T_j$ - the period of the periodical flow change of the spectrum detail ;
- $I_{id}^{(j)}$ - the average flow intensity corresponding to this detail;
- $\Omega_i^{(j)}$ - the radio source frequency corresponding to this detail.

*3.2 Results and analysis of the data processing*

In order to search for the periodical components in the signals of space masers the following radio sources were investigated in Pushchino RAO with the help of RT-22 radio telescope: W49N (9 sessions), W33B (3 sessions), Cep A (9 sessions), S128 (1 session), W3(2) (1 session), W44C (1 session), W75N (3 sessions), RT Vir (1 session). Together with these measurements performed in 2008 and aimed at the research discussed here, the previous measurements performed earlier in 2005-2007 in order to investigate the ultra-rapid variations of the flow for the same sources were also processed and analyzed (20 session more). The flow time dependencies for the various details of the spectra vary for different sources and dates. The results of the signals processing presented on Figs.2-5 show that for some details the flow does not reveal any essential changes (Fig.2), while some adjacent details reveal the compensating behavior, Fig.4, that has already been reported earlier in [6 - 9].

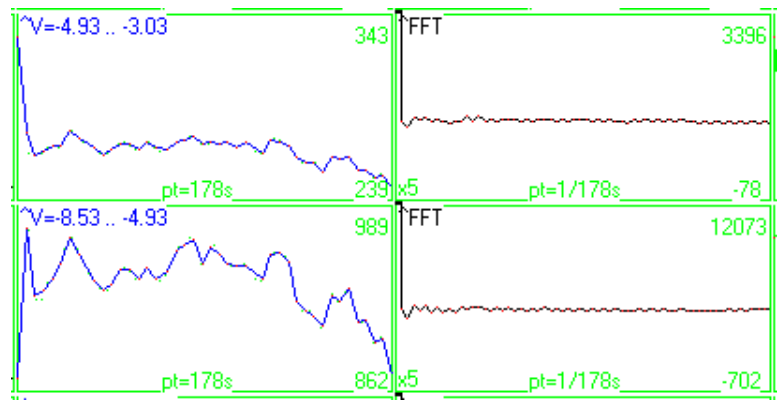

Fig. 2: W49N time dependencies and Fourier spectra, RT-22, Pushchino RAO, 12.02.2008

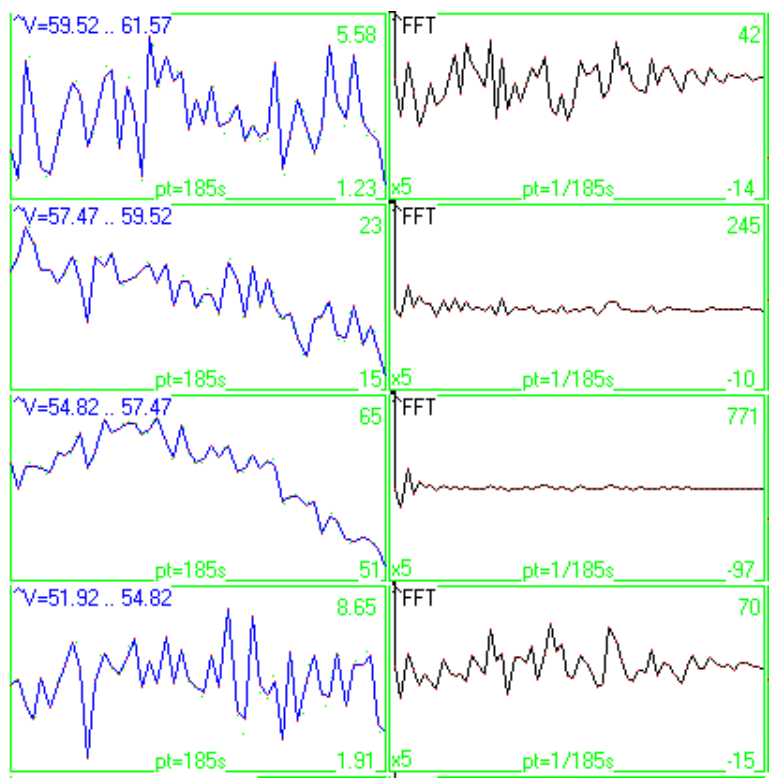

Fig. 3: W33B time dependencies and Fourier spectra, RT-22, Pushchino RAO, 26.06.2008

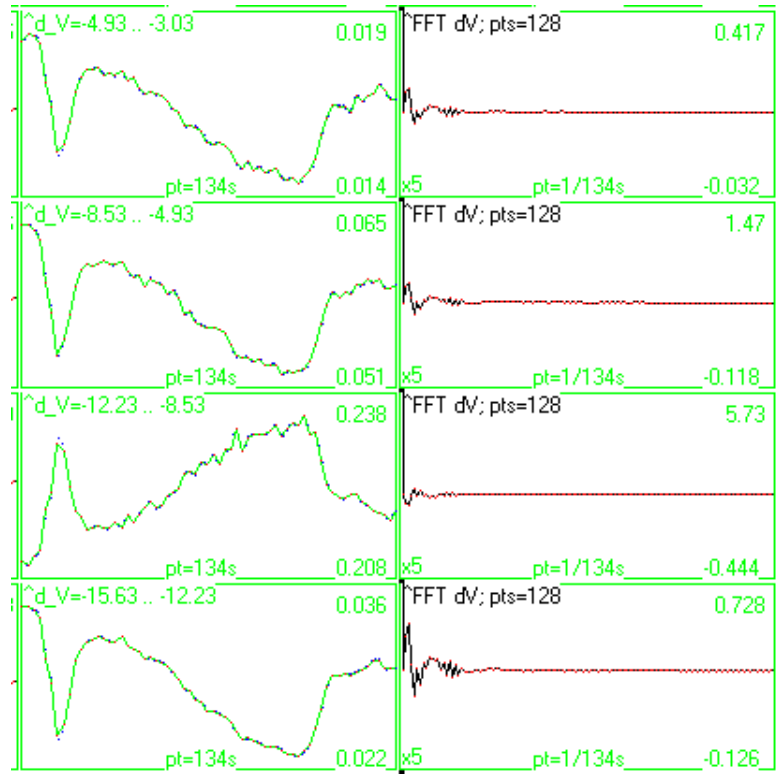

Fig. 4: W49N time dependencies and Fourier spectra with compensatory behavior, RT-22, Pushchino RAO, 11.02.2008

Notice, that such compensating behavior resembles the consequence of the OMPR theory [10] according to which there is a periodical cophased change of the intensity in the two sidebands of the central peak. This change is accompanied by the corresponding opposite changes of the intensity of the central peak – this shows that the OMPR signal is not due to the very small energy of the GW but is a result of the redistribution of the maser EMW energy in the conditions of the parametric resonance. As can be seen on Fig.4, the profile of the measured time dependence is more complicated, therefore, these plots can not be immediately interpreted as a result of the GW action on the maser, but this feature seems also important.

But in some data samples there were found the spectra details that demonstrated almost monochromatic periodic intensity changes. In view of the goal of this research, one should make sure that one and the same source reveals the varying component with the same period at different dates. On this stage of the research, this was found (Fig.5) for the Cep A radio source which is at the distance d = (0.70 ± 0.04) kpc from the Earth and has the coordinates RaJ $22^h 54^m 19.2^s$, Dec $61^0 45' 44.1''$. The corresponding dates of the observations were 07.02.08 and 24.06.08 and the periods of the intensity change were 22.2 min and 23.7 min. The difference in these two values could be explained by the different exposure times for two sessions (4 min and 7 min respectively). On the other dates this peak can be also seen on the plot, but alongside with it there are others that correspond to the other terms of the Fourier expansion, Fig.6.

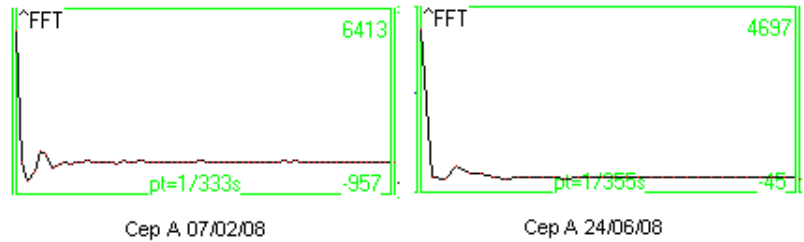

Fig. 5: Cep A time dependencies and Fourier spectra, RT-22, Pushchino RAO, 07.02.2008 and 24.06.2008

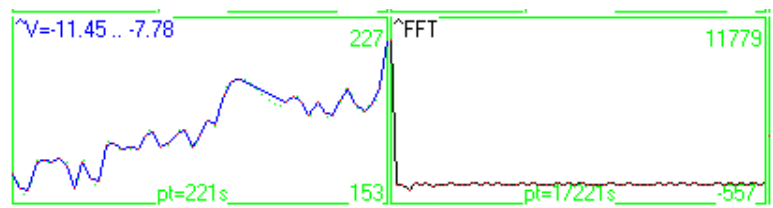

Fig. 6: Cep A time dependencies and Fourier spectra, RT-22, Pushchino RAO, 25.06.2008

It could be also seen that the high amplitude periodical components are present in the other sources signals too, but to make the results more reliable the exposure times should be diminished. This will be done in the observations to be performed soon with the modernized instrument.

*3.3 Interpretation*

The measured and analyzed variations of the radio sources' spectra intensities and their details show the existence of the periodical components that could be interpreted as the evidence of the possible GW action upon a space maser. Nowadays, no mechanism but the OMPR effect is suggested to explain the discovered periodicity in the flow variations presented on Fig.5. Though additional and more precise observations are still needed, the obtained results can be used for preliminary conclusions.

With regard to eq.(1), $v_j = 2D$, i.e. the period of the gravitational wave, $T_{GW}$, can be found $T_{GW} = 2T_j$. The known compact binaries with the periods corresponding to that of the Cep A periodic component are the double pulsars XTE J1807-294 with the period of 40.1 min [13], XTE J1751-305 with the period of 42.4 min [14] and XTE J0920-314 with the period of 44 min [14] and the double star V429 Car 10:41:17.5 - 59:40:37 which according to the measurements by Hipparcos [15] has a period of 44.64 min. Two last objects though having periods closer to the measured values do not satisfy the geometrical conditions (see Appendix). That is: if the distances from the GW source to the maser and from the GW source to the Earth are approximately equal and are essentially larger than the distance between the maser and the Earth, then the direction from the Earth to the GW source must be almost perpendicular to the line connecting the maser and the Earth.

The first two mentioned binaries more or less suffice the geometrical conditions but their periods are less than the measured ones. Thus, the GW action of the known binaries could lead to the appearance of the discovered periodic component on the Cep A maser spectrum if the forthcoming more precise measurements show that its period is a couple of minutes less. If it is not so, the search for other binaries is needed.

## 4. Discussion

The main result of this paper is the discovery of the periodical component in the spectrum of the space maser. Since there is no other mechanism than OMPR which is able to explain the periodical character of the detail's intensity change, we can assume that this result could be caused by the action of the periodical GW on the space maser. The obtained data makes it possible to suggest that the radio source (space maser) Cep A is affected by the gravity radiation emitted by the compact binary system which could be specified in the forthcoming research. The used observation and processing methods are unable to detect the periodical changes with the periods less than the doubled exposure time. That is why on the next stage of the observations the set up will be modernized and the exposure time will be diminished to tens of seconds. This will not only provide an order increase in accuracy but will make it possible to involve such GW sources as long period pulsars whose periods are known with high accuracy.

The registration of such a signal can give a new experimental evidence of the GW existence and since the value of the OMPR signal is comparable to that of the regular maser peak, such observations could lead to the design of a GW map of the sky. The independent observations are obviously needed to confirm our result.

## 5. Acknowledgment

This work was supported by the RFBR grant No. 08-02-01179-a.

## 6. Appendix

**Brief theory of the OMPR effect**

Let us regard a two-level atom in the strong monochromatic quasi-resonant electromagnetic field. The system of Bloch's equations for the components of the density matrix components is

$$\frac{d}{dt}\rho_{22} = -\gamma\rho_{22} + 2i\alpha_1 \cos(\Omega_1 t - k_1 y)(\rho_{21} - \rho_{12})$$
$$[\frac{\partial}{\partial t} + v\frac{\partial}{\partial y}]\rho_{12} = -(\gamma_{12} + i\omega)\rho_{12} - 2i\alpha_1 \cos(\Omega_1 t - k_1 y)(\rho_{22} - \rho_{11}) \quad (A1)$$
$$\rho_{22} + \rho_{11} = 1$$

Here $\rho_{22}$, $\rho_{11}$ are the populations of the levels, $\rho_{12}$, $\rho_{21}$ are the polarization terms, $\gamma$ and $\gamma_{12}$ are the longitudinal and transversal decay rates of the atom (if level 1 is the ground level, $\gamma_{12} = \gamma/2$); $k$ is the $y$-component of the wave vector of the EMW, $v$ is the atom velocity along the $Oy$-axis pointing at the detector, $\gamma << \alpha_1$ is the condition of the strong field. The dynamics of this system in cases when various parametric resonances are possible was investigated in [10].

Let this atom belong to a saturated space maser in the field of the periodic GW emitted by a compact binary star or by a pulsar and propagating anti-parallel to the $Ox$-axis pointing at the GW-source. The GW acts on the atomic levels, on the maser radiation and on the geometrical location of the atom. In [11] it was shown that the first effect is much smaller than the other two effects. The action of the GW on the monochromatic EMW could be accounted for by the solution of the eikonal equation,

$$g^{ik}\frac{\partial \psi}{\partial x^i}\frac{\partial \psi}{\partial x^k}=0, \quad (A2)$$

in the gravitational field for the empty space (far from masses)

$$g^{ik}=\begin{pmatrix} 1 & 0 & 0 & 0 \\ 0 & -1 & 0 & 0 \\ 0 & 0 & -1+h\cos\frac{D}{c}(x^0-x^1) & 0 \\ 0 & 0 & 0 & -1-h\cos\frac{D}{c}(x^0-x^1) \end{pmatrix} \quad (A3)$$

The atom velocity, $v$, could be obtained from the solution of the geodesic equation

$$\frac{d^2 x^i}{ds^2}+\Gamma^i{}_{kl}\frac{dx^k}{ds}\frac{dx^l}{ds}=0 \quad (A4)$$

(and not from the solution of the geodesic declination equation as in the calculations of the displacements of the parts of the laboratory setups, designed for the detection of the GW). The solution of eq.(A4) gives [11] the periodic dependence of velocity on time

$$v=v_0+v_1\cos Dt; v_1=hc \quad (A5)$$

The solution of eq.(A2) with regard to eq.(A3) shows that the action of the GW causes a phase modulation of the EMW. Since $h$ is very small, the phase modulated EMW can be presented [12] as a superposition of amplitude modulated EMWs. Substituting this result together with eq.(A5) into the system eq.(A1), one can see that the parametric resonance is possible provided the following conditions are fulfilled:

amplitude conditions:

$$\frac{\gamma}{\alpha_1}=\Gamma\varepsilon; \Gamma=O(1); \varepsilon<<1 \quad (A6)$$

- the space maser is a saturated one (notice that for $H_2O$-maser $\gamma \sim 10^{-14} s^{-1}$, i.e. it corresponds to the metastable state);

$$\frac{\alpha_2}{\alpha_1}=\frac{\omega h}{4D}=a\varepsilon; a=O(1); \varepsilon<<1, \quad (A7)$$

$$\frac{kv_1}{\alpha_1}=\frac{\omega h}{\alpha_1}=\kappa\varepsilon; \kappa=O(1); \varepsilon<<1 \quad (A8)$$

- relations for the distance between the GW-source and maser, source and maser characteristics and the frequencies ratio (the amplitude of the GW can be defined by

$$h = \frac{GMR^2 D^2 g_e}{c^4 r_S});$$

frequency condition:
$$(\omega - \Omega_1 + kv_0)^2 + 4\alpha_1^2 = D^2 + O(\varepsilon) \quad \Rightarrow D \sim 2\alpha_1 \quad (A9)$$

The solution is performed by the asymptotic expansion method, the small parameter being $\varepsilon = \gamma/\alpha_1$ (notice, that $\alpha_2/\alpha_1 = \varepsilon$ has the same small parameter value). There is also the geometrical condition: if the Earth and the GW source are located on the ends of the diameter, then the maser must be close to the circumference built on this diameter.

The analysis given in [12] shows that these conditions can be sufficed in natural environment, the examples of the suitable astrophysical systems can be also found in [12]. If the conditions eqs.(A6-A9) are fulfilled, then the principal term of the asymptotic expansion for $Im(\rho_{21})$ which characterizes the scattered radiation energy flow can be calculated explicitly. The effect of the OMPR is that at the frequency shifted by $D$ from the central peak of the EMW (that is from the usual signal of the space maser), the energy flow is proportional to $\varepsilon^0$, i.e. has the zero order in the powers of the small parameter of the expansion, and has the form eq.(1). It means that the energy flow at this frequency is periodically amplified and attenuated with the (doubled) frequency of the GW.

## References

[1] Gladyshev, V. 2000, Irreversible electromagnetic processes in the astrophysical problems. Moscow, Bauman University.

[2] Bochkarev N.G. 1992, Foundations of the interstellar medium physics. Moscow, MGU.

[3] Zel'dovich Ya.B., Novikov I.D. 1988, Physics of the interstellar medium. Moscow, Nauka.

[4] Elitzur, M. 1992, Astronomical masers. Boston, Kluwer Academic Publishers.

[5] Cook, A.H. 1977, Celestial masers. London, Cambridge University Press.

[6] Samodurov, V.A. and Logvinenko S.V. 2001, Astron. Rep. ,Vol. 45, p. 339.

[7] Richards, A. M. S. et al. 2005, Astrophysics and Space Science, Vol. 295, p. 19.

[8] Samodurov, V. A. et al. 2006, Astron. and Astrophys. Transactions, Vol. 25, No.5, p. 393.

[9] Samodurov, V. A. et al. 2008, Proc. Int. Conf. "Astronomy and Astophysics of the beginning of the XXI c.", Moscow, Sternberg Inst., p.75.

[10] Siparov, S.V. 1997, Phys.Rev.A, Vol. 55, p. 3704; Kazakov, A.Ya,, Siparov, S.V. 1997, Opt. Spektrosk. Vol. 83, p. 961 (rus); Siparov, S.V. 1998, J.Phys.B, Vol. 31, p. 415.

[11] Siparov, S. 2004, Astronomy&Astrophysics, Vol.416, p.815.

[12] Siparov, S. 2006, Proc.Conf.PIRT-06, London; Siparov, S. 2006, Hyper-complex numbers in Geometry and Physics, Vol.6, p.155; Siparov, S. 2007, in Space-Time Structure, Algebra and Geometry, Moscow, p.495; Siparov, S. 2007, Proc. Conf. PIRT-07, p.282, Moscow; Siparov, S. 2008, AMAPN, 24(1), p.135.

[13] Markwardt C.B., Juda, M. and Swank J.H., 2003, IAU Circ., Vol.2. p. 8095.

[14] Markwardt, C.B. 2002, Astrophys.J., Vol. 576, p. L137.

[15] Hipparcos No.52308, Hipparcos Variability Annex: Tables. Part2: Unsolved Variables Volume: 11[th], 1997.